\documentclass[conference]{IEEEtran}
\IEEEoverridecommandlockouts
\usepackage{graphicx}
\usepackage{color}
\usepackage[utf8]{inputenc}
\usepackage{amsmath,amssymb,amsfonts,amsthm}
\usepackage{algorithmic}
\usepackage{multirow}

\title{Low-Power Data Streaming in Systolic Arrays with Bus-Invert Coding and Zero-Value Clock Gating}
\author{
\IEEEauthorblockN{Christodoulos Peltekis, Dionysios Filippas, Giorgos Dimitrakopoulos
\thanks{This work was supported by a research grant of Siemens EDA to Democritus University of Thrace for "HLS Research for Systems-on-Chip".}} 
\IEEEauthorblockA{Electrical and Computer Engineering\\ 
Democritus University of Thrace, Xanthi, Greece}
\and
\IEEEauthorblockN{Chrysostomos Nicopoulos}
\IEEEauthorblockA{Electrical and Computer Engineering\\  University of Cyprus, Cyprus}}

\begin{document}

\maketitle

\begin{abstract}
Systolic Array (SA) architectures are well suited for accelerating matrix multiplications through the use of a pipelined array of Processing Elements (PEs) communicating with local connections and pre-orchestrated data movements. Even though most of the dynamic power consumption in SAs is due to multiplications and additions, pipelined data movement within the SA constitutes an additional important contributor. The goal of this work is to reduce the dynamic power consumption associated with the feeding of data to the SA, by synergistically applying bus-invert coding and zero-value clock gating. By exploiting salient attributes of state-of-the-art CNNs, such as the value distribution of the weights, the proposed SA applies appropriate encoding only to the data that exhibits high switching activity. Similarly, when one of the inputs is zero, unnecessary operations are entirely skipped. This selectively targeted, application-aware encoding approach is demonstrated to reduce the dynamic power consumption of data streaming in CNN applications using Bfloat16 arithmetic by 1\%--19\%. This translates to an overall dynamic power reduction of 6.2\%--9.4\%.  
\end{abstract}

\section{Introduction}
Matrix multiplications are at the heart of deep learning algorithms and -- in hardware -- they naturally map onto Systolic Arrays (SA)~\cite{why-systolic}. SAs have a long history of wide applicability~\cite{why-systolic}, while, recently, their design has regained interest due to the large volume of rapidly emerging machine learning applications. Tensor Processing Units (TPUs)~\cite{tpu} and other related architectures~\cite{scalesim, auto-sa, meissa, factored-sa} are characteristic examples of newly designed SA architectures/derivatives.

\begin{figure}
\centering
\includegraphics[width=0.9\columnwidth]{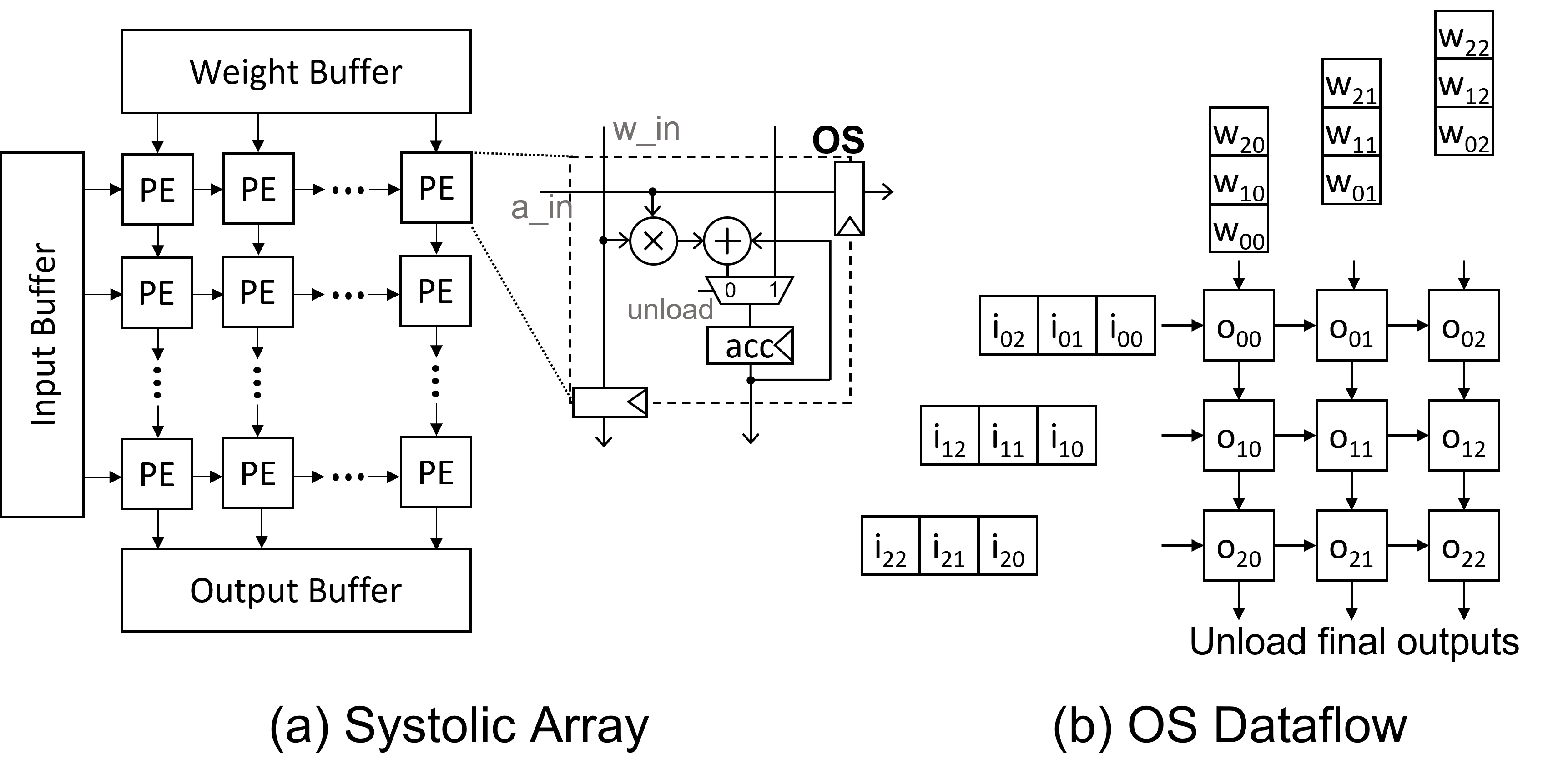}
\caption{An output-stationary SA and its corresponding dataflow.}
\label{f:sa-baseline}
\end{figure}

The typical SA hardware structure consists of an array of Processing Elements (PEs), as depicted in Fig.~\ref{f:sa-baseline}(a). Each PE consists of a multiplier and an adder and necessary registers to appropriately pipeline the streaming operation. The SA is fed by local memory banks placed on the West and North edges of the array, while the output results are collected on the South edge. When the sizes of the matrices are larger than the size of the SA, matrix multiplication is executed in tiles, where the size of each tile (i.e., sub-matrix) matches the size of the SA.

The \textit{dataflow} type employed by the SA determines the internal structure of the PEs and how the matrix multiplication, $A\times B$, is executed~\cite{arrayflex}. 
For instance, in \textit{output-stationary} dataflow, the matrices $A$ and $B$ arrive synchronously from the West and North borders of the SA, respectively, and the result is accumulated locally within each PE. The micro-architecture of each PE for the output-stationary dataflow type is shown in Fig.~\ref{f:sa-baseline}(a), while the actual data flow is visualized in Fig.~\ref{f:sa-baseline}(b). The unloading of the final result occurs separately, upon completion of the computation phase. To increase applicability, flexibility, and PE utilization, configurable SAs can support various dataflow types~\cite{hetero-sa, dataflow-mirroring, arrayflex}. 

The focus of this work is on the reduction of dynamic power consumption within the SA, which is directly proportional to the amount of activity flowing through the system. We explore for the first time -- to the best of our knowledge -- the synergistic application of established data-coding techniques to reduce the dynamic power consumption involved in the data-and-weight loading process. The goal is to \textit{optimally} apply data coding using a \textit{targeted} approach that minimizes data switching activity in a cost-effective manner. Overall, the contributions of this work can be summarized as follows:
\begin{itemize}
  \item
  We explore the value distribution of the \textit{weights} of modern CNNs to identify the data segments within the weights that exhibit high switching activity. This information allows for optimal application of encoding; data segments that exhibit low switching activity are simply not encoded. Hence, the data switching activity is minimized at the minimum possible hardware cost.
  \item The statistical analysis of the weight values constitutes a first attempt to identify and exploit useful numerical attributes in the new and increasingly prevalent floating-point arithmetic formats, such as Bfloat16~\cite{bfloat-def, aicas}. Reduced precision floating-point arithmetic is rapidly gaining traction in deep learning, where sufficient inference accuracy is not always achieved using only integers~\cite{ten-lessons, tensorfloat}. 
  \item Zero-value clock gating is selectively applied only to the \textit{inputs} of the CNN layers to exploit the presence of zero values generated by the Rectified Linear Unit (ReLU) activation function. Unnecessary operations involving zeros are skipped altogether within the SA.
\end{itemize}


Extensive evaluations using complete CNN applications driven with test images from the ImageNet database~\cite{imagenet} demonstrate the efficacy and efficiency of the proposed architecture. The switching activity is reduced by 29\%, on average, as compared to SAs with unencoded data values, which translates into total dynamic power savings of 9.4\%, on average, in ASIC hardware implementations.

\section{Dynamic Power Consumption in SAs}

Overall, the dynamic power consumption of a systolic array when computing a matrix multiplication consists of three main components:

{\bf Data and weight loading:} This involves the loading of data and weights in the horizontal and vertical directions, respectively. In output-stationary dataflow, the data loading occurs in parallel to the actual computation, from the West and the North side of the SA, in an orchestrated manner. The cost of data loading involves the power expended in the clocked elements (pipeline registers) and the wires that connect them, and it is directly related to the switching-activity profile of the incoming data.

{\bf Computation power:} The dynamic power consumed for the actual computation, i.e., multiplication and addition. This part is the major power consumer and depends mainly on the size and the arithmetic format of the processed data and the complexity of the datapath logic. Quantization for Deep Neural Networks (DNNs) retains the inference-time model quality using integer-only arithmetic. State-of-the-art Machine Learning (ML) models, such as transformers, still require floating-point arithmetic, even if this refers to low-precision 8-/16-bit formats~\cite{ten-lessons}.

{\bf Sum accumulation/unloading:} The power cost of moving the partial or final sums. In output-stationary dataflow, this occurs only once, at the end of the computation phase. The dynamic power associated with the movement of partial sums involves the power consumed in pipeline registers and the wires that connect them.

This work tackles the first of the aforementioned components; namely, the consumption attributed to the loading of the data and weights. The subsequent experimental evaluation will demonstrate that, even though we do not focus on the major consumer of power (i.e., computation), the reaped overall savings are still quite substantial.

\section{Low-Power Data Loading and Movement}

To decrease the dynamic power consumption due to data movement, the ultimate goal is to minimize the switching activity within the horizontal and vertical register pipelines of the array. While there is a wealth of proposed approaches in the literature to lower switching activity, this work is the first -- to the best of our knowledge -- to apply such techniques within the SA context and quantitatively assess their effectiveness when executing real CNN applications.

\subsection{Reduction of switching activity}
The three most relevant techniques are: (a) data-driven clock gating, (b) bus-invert coding, and (c) zero-value clock gating.

The first one is the most fine-grained mechanism for clock gating, since it is deployed in flip-flops at the gate level. The clock signal driving a flip-flop is disabled (gated) if the flip-flop state would remain unchanged in the next clock cycle~\cite{data-driven-clock-gating}. To reduce overhead, several flip-flops may be grouped together and driven by the same clock gating logic generated by ORing the enabling signals of the individual flip-flops. Of course, this coarser granularity tends to lower the disabling effectiveness. Therefore, it is beneficial to group flip-flops whose switching activities are highly correlated. This constraint renders this approach infeasible for CNN applications, because it is very difficult to find groups of bits that remain the same. Consequently, the aggressive bit-level application of this technique would incur undue overhead, whereas the coarser group-level implementation would not be effective for CNNs.

Hence, we focus on the remaining two techniques that are well suited for a systolic array setting.

{\bf (1) Bus-Invert Coding (BIC)}: This well-known technique~\cite{bic} lowers the bus activity by reducing the number of bit transitions. The algorithm computes the hamming distance between the present and next bus values, i.e., the number of bits in which they differ. If this number is more than half of the bus width, then the \textit{complement} of the next datum is transmitted instead. Alongside the transmitted data, the BIC scheme also transmits a single bit (`inv-bit'), indicating whether the transmitted data is inverted or not. A variant of the BIC technique, called {\bf Segmented Bus-Invert Coding}, can be applied to different segments of the bus separately~\cite{segmented-bic}. For instance, BIC could be applied to the mantissa (fraction) and the exponent fields of a floating-point number \textit{independently}. Other more elaborate coding techniques that are useful when driving large buses are not considered~\cite{encode-wide-busses, spatial-encoding}, since their encoding/decoding overhead diminishes any savings in switching activity at the PE level.

{\bf (2) Zero-Value Clock Gating}: This technique can be used when the input or/and the weight have zero value. Since the result of any multiplication involving at least one zero value is zero, there is no need to perform the multiplication. Therefore, upon detecting a zero input value, the pipeline is `frozen' (amounting to inserting a bubble) and an `is-zero' bit is asserted. This bit is subsequently used to clock-gate the registers and to bypass the result of the multiplication, which is known a priori to be equal to zero.

\subsection{Systolic Arrays with Selective Coding and Clock Gating}
To cost-effectively apply the above two techniques within an SA, it is imperative to identify, and account for, some nuances of CNNs, which constitute our primary target applications. 

\begin{figure}
    \begin{tabular}{c}
    \includegraphics[width=0.95\columnwidth]{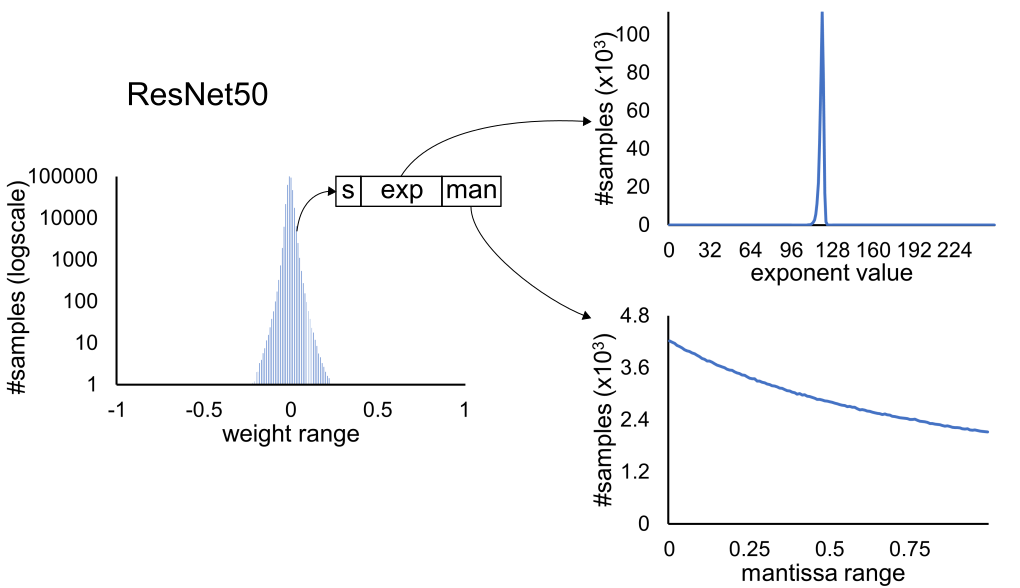}\\
    \\
    \includegraphics[width=0.95\columnwidth]{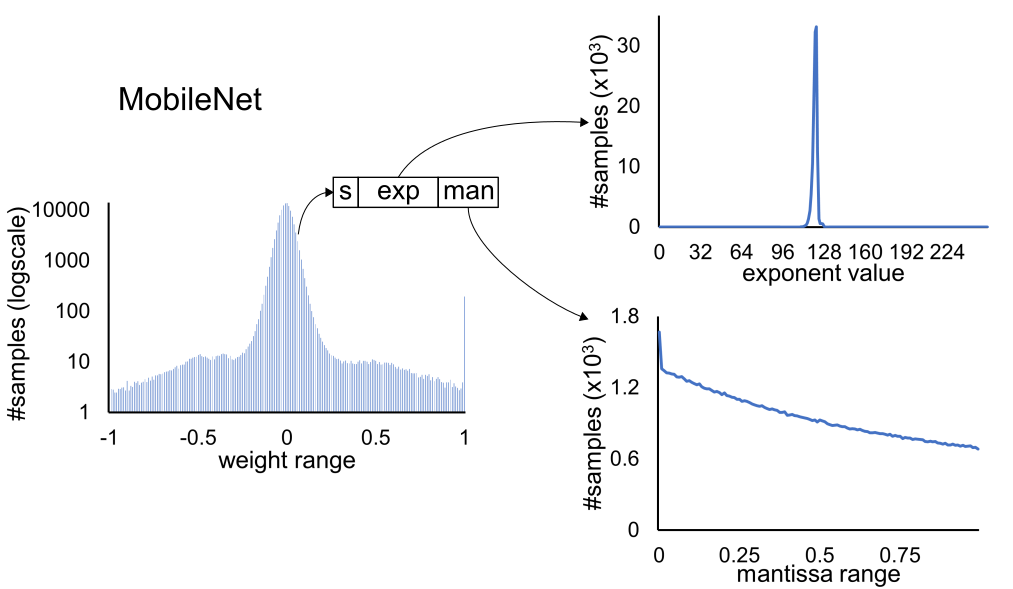}
    \end{tabular}
    \caption{The observed value distributions of the weights of all layers of the ResNet50~\cite{resnet} and MobileNet~\cite{mobilenet} CNNs. The exponent values of the corresponding Bfloat16 numbers are highly concentrated, while the mantissa values are almost uniformly distributed in the entire available dynamic range.}
    \label{f:weight-val-distribution}
\end{figure}

For instance, in CNNs, the weights of each layer are determined during training and remain unchanged throughout the inference phase. The weights encountered in CNNs have some interesting attributes that can be exploited by the BIC technique. In an attempt to identify useful -- in terms of encoding potential -- numerical attributes exhibited by the weights of modern CNN applications, we perform a detailed statistical analysis of the observed value distributions when using the widely utilized Bfloat16 floating-point format. Fig.~\ref{f:weight-val-distribution} illustrates the value distributions of the weights of all layers of  two state-of-the-art CNNs: ResNet50~\cite{resnet} and MobileNet~\cite{mobilenet}. Specifically, the histogram of the weight values is further analyzed into the distributions of the \textit{exponent} and \textit{mantissa} values. These weight values are bounded to the  range $[-1,1]$ from the training step.

As shown in Fig.~\ref{f:weight-val-distribution}, the weight values are highly concentrated around zero, i.e., their absolute values are very small. As a result, when the weights are represented as Bfloat16 numbers, their exponent values are highly concentrated close to the bias value. This high concentration implies that consecutive exponent values have very few bits that are different, thereby rendering the BIC technique \emph{non-beneficial for the exponents}. On the other hand, the values of the \textit{mantissa} field of the CNN weights are almost uniformly distributed, thus making them amenable to BIC. Hence, for the \textit{weights}, we use BIC targeting \textit{only the mantissa} field.

On the other hand, the input data is different for each input image and is highly dependent on the selected activation function. Therefore, there is no clear statistical attribute that can be exploited for the inputs, other than the abundant zero values generated by the ReLU activation function in each layer. Hence, the zero-value clock gating technique need \textit{only be applied to the inputs} of the CNN layers. This ensures that no power is wasted on redundant operations. Note that the abundance of zeros can be artificially increased in the weights, too, by enabling weight pruning techniques. However, such approaches are out of the scope of this work.

\begin{figure}[t]
\centering
\includegraphics[width=0.9\columnwidth]{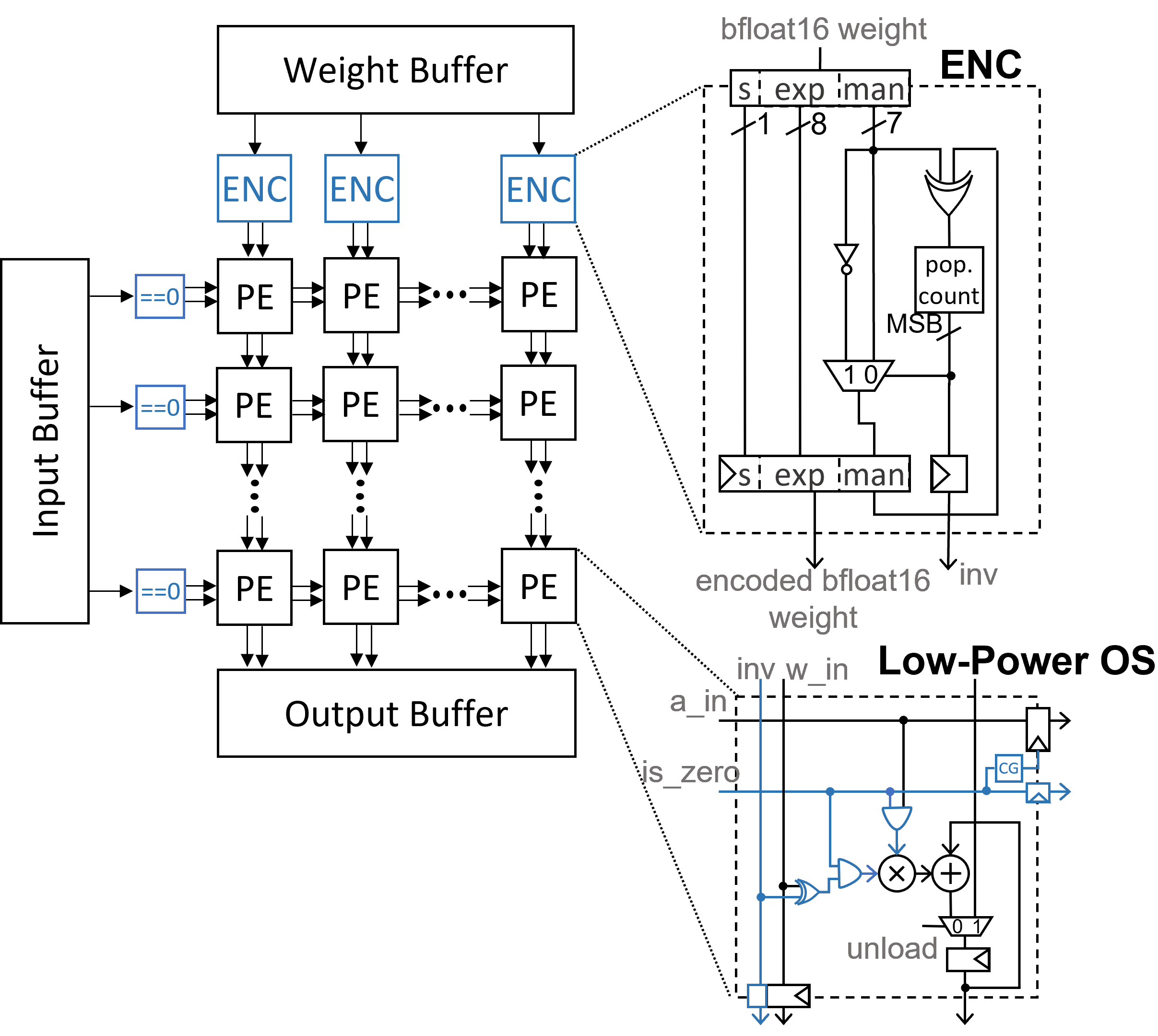}
\caption{The proposed low-power SA architecture that combines two power-saving techniques to reduce the switching activity as data flows through the array of PEs. The inputs pass through zero-detection logic at the West edge of the SA. The weights are encoded at the North edge, prior to entering the array, and the actual values are recovered within each PE for the ensuing calculations.}

\label{f:low-power-sa}
\end{figure}

The proposed SA architecture, which effectively combines the BIC and zero-value clock gating techniques, is shown in Fig.~\ref{f:low-power-sa}. As compared to a baseline SA (i.e., without power-saving functionality), the additional new logic -- shown in color blue -- is primarily found on the North and West edges of the SA within the \textit{Encoding} (`ENC') modules and the zero-value checkers, respectively. As illustrated in Fig.~\ref{f:low-power-sa}, each Encoding module implements the BIC technique only on the mantissas of the weights, as previously explained. This targeted/selective BIC approach reduces the incurred area overhead and avoids redundant power consumption. In the proposed architecture, there is also some lightweight new logic within each PE. Specifically, XOR gates are added to recover the original value of the mantissa of each weight, if it was inverted by the BIC technique. Clock-gating logic is also added in the input datapath, which is triggered by the `is-zero' bit. The latter is also used to data-gate the multiplier, in order to eliminate wasted power in multiplications with zero.

It should be noted that the zero-value clock gating mechanism has been applied to SAs in the past~\cite{zero-gating}. However, in this work, we use it in \textit{conjunction} with BIC to reap the extra benefits of the \textit{synergistic} effect of both techniques operating in tandem. More importantly, the employed \textit{targeted} encoding approach selectively applies the most appropriate scheme to each data type (inputs vs. weights) to reap the most power savings with the minimum hardware cost.

\begin{figure*}
    \centering
    \includegraphics[width=0.99\textwidth]{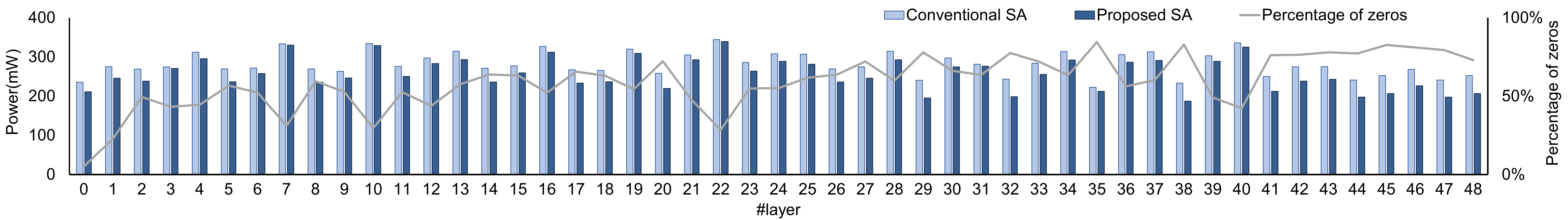}
    \caption{The per-layer power consumption and the percentage of zero-value inputs for the ResNet50~\cite{resnet} CNN.}
    \label{f:power-resnet}
\end{figure*}

\begin{figure}
    \centering
    \includegraphics[width=\columnwidth]{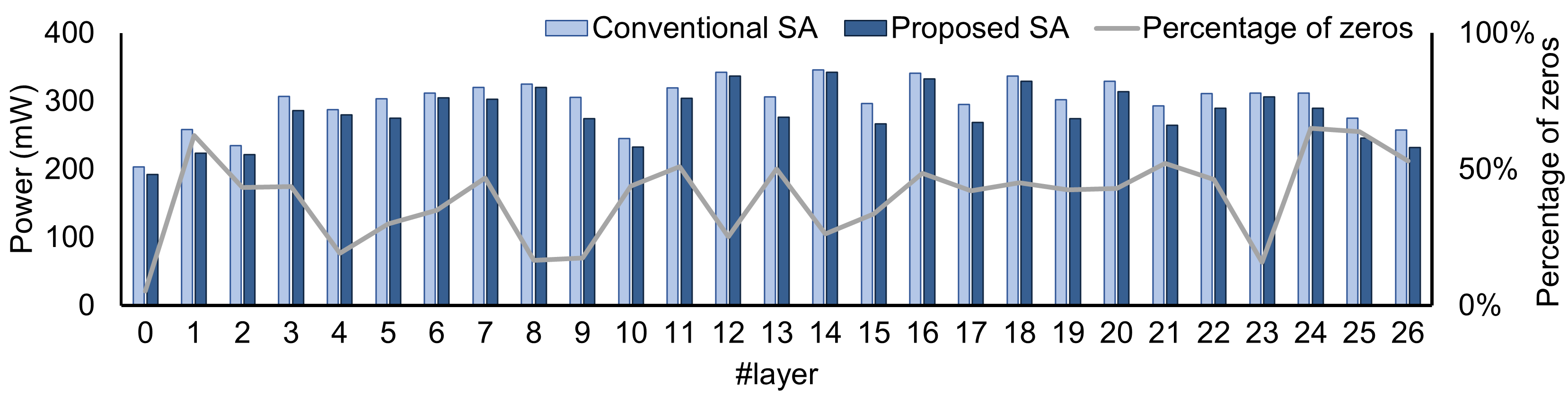}
    \caption{The per-layer power consumption and the percentage of zero-value inputs for the MobileNet~\cite{mobilenet} CNN.}
    \label{f:power-mobilenet}
\end{figure}

\section{Experimental results}

In this section, we demonstrate the effectiveness of the proposed SA architecture in reducing the data switching activity and, consequently, the overall power consumption. The new design, which selectively applies the BIC and zero-value clock gating techniques, is compared to a conventional SA, i.e., one with no power-saving features.

Both architectures under comparison were implemented in C++ and synthesized to Verilog RTL using Catapult HLS driven by a commercial-grade 45-nm standard-cell library. 
Both SA designs have an array size of 16$\times$16 PEs.
Bfloat16 multiply and add operators were implemented using Catapult's built in floating point datatypes. Final timing/area results are derived from the Oasys logic synthesis tool.
Power was estimated after synthesis using the PowerPro power analysis and optimization tool.

The hardware area overhead incurred by the extra logic in the proposed design is 5.7\%. It is important to note that this percentage overhead will, in fact, decrease with larger SA sizes, because the number of additional encoders scales \textit{linearly} with the size of the SA, whereas the number of additional PEs scales \textit{quadratically}.

In any case, the small area overhead is fully amortized by the larger reaped power savings, as illustrated in Figs.~\ref{f:power-resnet} and~\ref{f:power-mobilenet}. These figures report the per-layer power consumption in two state-of-the-art CNNs: ResNet50~\cite{resnet} and MobileNet~\cite{mobilenet}. The percentage of input values that are zeros is also included for each layer. The power consumption numbers and the zero-value percentages are the average of 100 randomly picked images from the ImageNet database~\cite{imagenet}. As shown in Figs.~\ref{f:power-resnet} and~\ref{f:power-mobilenet}, the proposed architecture yields 1\%--19\% per-layer power savings, as compared to the conventional SA. These per-layer savings translate to an \textit{overall power reduction} of 9.4\% for ResNet50~\cite{resnet} and 6.2\% for MobileNet~\cite{mobilenet}.

It can be observed in Figs.~\ref{f:power-resnet} and~\ref{f:power-mobilenet} that, in the layers where the percentage of zero-value inputs is high, the power consumption in the proposed design is -- in most cases -- much lower than in the conventional SA, due to the extensive use of the zero-value clock gating technique. Nevertheless, when the number of zero values becomes very high, there are more cases of multiple consecutive zero inputs. Naturally, these cases also benefit the power consumption of the conventional SA design. Overall, the synergistic application of the BIC and zero-value clock gating techniques in the proposed architecture yields consistent power savings in all the layers of the two CNNs.

\section{Conclusions}

Even though data- and weight-loading constitute only a portion of the total power consumed in a systolic array, a reduction in switching activity while traversing the array still yields significant overall power savings. The proposed SA architecture reduces the overall power consumption by appropriately encoding the data flowing through the array to reduce the switching activity. By exploiting key attributes of modern CNNs, such as the value distribution of the weights and the prevalence of zero values in the inputs, the new design applies the most appropriate scheme to each data type to maximize the power savings with the minimum hardware cost. This targeted, application-aware encoding yields overall power savings of 6.2\%--9.4\% in state-of-the-art CNN applications.

\bibliographystyle{IEEEtran}
\bibliography{refs}

\end{document}